# Securing the Invisible Thread: A Comprehensive Analysis of BLE Tracker Security in Apple AirTags and Samsung SmartTags

Hosam Alamleh, Michael Gogarty, David Ruddell, and Ali Abdullah S. AlQahtani, *Member, IEEE*

*Abstract*—This study presents an in-depth analysis of the security landscape in Bluetooth Low Energy (BLE) tracking systems, with a particular emphasis on Apple AirTags and Samsung SmartTags, including their cryptographic frameworks. Our investigation traverses a wide spectrum of attack vectors such as physical tampering, firmware exploitation, signal spoofing, eavesdropping, jamming, app security flaws, Bluetooth security weaknesses, location spoofing, threats to owner devices, and cloud-related vulnerabilities. Moreover, we delve into the security implications of the cryptographic methods utilized in these systems. Our findings reveal that while BLE trackers like AirTags and SmartTags offer substantial utility, they also pose significant security risks. Notably, Apple's approach, which prioritizes user privacy by removing intermediaries, inadvertently leads to device authentication challenges, evidenced by successful AirTag spoofing instances. Conversely, Samsung SmartTags, designed to thwart beacon spoofing, raise critical concerns about cloud security and user privacy. Our analysis also highlights the constraints faced by these devices due to their design focus on battery life conservation, particularly the absence of secure boot processes, which leaves them susceptible to OS modification and a range of potential attacks. The paper concludes with insights into the anticipated evolution of these tracking systems. We predict that future enhancements will likely focus on bolstering security features, especially as these devices become increasingly integrated into the broader IoT ecosystem and face evolving privacy regulations. This shift is imperative to address the intricate balance between functionality and security in next-generation BLE tracking systems.

*Index Terms*—Bluetooth Low Energy, Security Analysis, Apple AirTags, Samsung SmartTags, Digital Authentication, Signal Spoofing, BLE.

## I. Introduction

Since its advent in 2010, Bluetooth Low Energy (BLE) technology has significantly evolved, catalyzing the development of a myriad of products, notably BLE trackers like Apple's AirTag and Samsung's Galaxy SmartTag [1]. These innovative devices assist users in locating misplaced or lost items by pairing with smartphones and leveraging a network of devices to relay location information when out of range, functionality encapsulated in the 'Lost Mode'.

Despite their widespread adoption and utility, there remains a conspicuous gap in the comprehensive examination of the security aspects of these BLE trackers. This gap is partly attributed to the multifaceted nature of potential attacks and the sensitivity of the information at risk. In light of this, our paper embarks on a detailed security analysis of BLE trackers, primarily focusing on Apple AirTags and Samsung Galaxy SmartTags. We aim to illuminate the spectrum of potential risks and vulnerabilities these devices harbor, thereby advancing the understanding of their security implications.

Our research is pivoted on several critical questions:

- **RQ1:** What methods could be employed to compromise BLE tracking systems, and what would be the repercussions of these attacks?
- **RQ2:** How does the BLE tracker designs by apple and samsung impact user privacy and security?
- **RQ3:** What features are anticipated to be incorporated into future BLE trackers to address existing challenges?

To address these queries, we delve into the intricacies of the communication protocols, firmware, and encryption mechanisms deployed in these trackers. Additionally, we assess the broader implications of these systems' implementations on user privacy and security.

In doing so, we contribute to bridging the existing research gap by providing a holistic view of the security landscape surrounding BLE trackers. This investigation is not only crucial for users and manufacturers but also pivotal for the broader field of network security, especially as these devices become increasingly integrated into the fabric of the Internet of Things (IoT).

The remainder of this paper is structured as follows: Section II provides a detailed overview of the operational framework of BLE tracking systems, focusing on the roles and interactions of various components such as BLE trackers, owner devices, helper devices, and cloud infrastructure. Section III reviews existing literature and research in the field. This section contextualizes our study within the broader scope of BLE tracker security, highlighting the contributions of previous studies and identifying the gaps our research aims to fill. Section IV presents an in-depth examination of the various potential attack vectors targeting BLE tracking systems. Section V interprets the findings from the previous sections, discussing the broader implications of our analysis for BLE tracker implementations, user privacy, and system security. This section also explores potential future trends and developments in BLE tracking technology. Section VI encapsulates the key insights and contributions of our study.

## II. PRELIMINARY

BLE trackers, as integral parts of tracking networks, are pivotal in aiding users to locate and manage their belongings or devices. These networks typically leverage a network of Bluetooth-enabled devices, such as trackers or tags, and a central platform or application facilitating communication. Notable examples include the Apple FindMy [2] network and the Samsung SmartThings [3] network. The BLE tracking framework comprises:

1) **BLE Trackers**: Portable devices with BLE technology for communicating with nearby smartphones, aiding users in finding lost or misplaced items.
2) **Owner's Device**: Paired with the BLE tracker, applications like FindMy for Apple AirTags and SmartThings for Samsung enable tracking.
3) **Helper Devices**: Engage in community-based tracking, reporting the location of BLE trackers in lost mode to the server.
4) **The Cloud Server**: The backend infrastructure supporting the BLE tracker system, processing and storing data, accessible via dedicated smartphone apps.

Figure 1 presents the BLE Tracking System Required Components. The BLE tracking framework follows a series of steps: activation of lost mode when the tracker moves out of the owner's device range, detection and transmission of the token by helper devices, and the subsequent forwarding of this information to the owner's device.

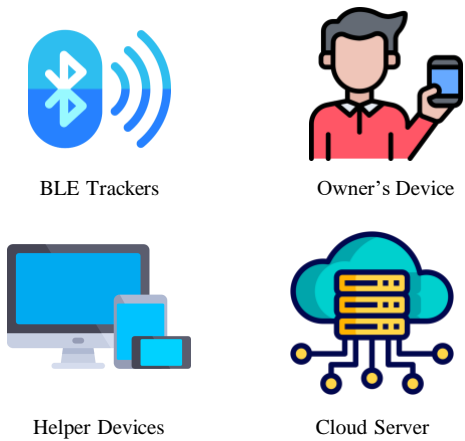

Fig. 1: Required Components; Icons From [4]

### A. Operational Differences Between AirTags and SmartTags

This section delves into the comparative analysis of Apple AirTags and Samsung SmartTags, focusing on their operational differences, and contextualizes these within the broader evolution of BLE trackers.

*1) Data Encryption and Decryption*
- **Apple AirTags:** utilize a public-key cryptography system (EC P-224), effectively balancing security with user privacy. The encryption process occurs at the helper device, while decryption is managed at the owner's device, as detailed in the following equations:

$$\text{Enc}_M = \text{Encrypt}_{\text{EC P-224}}(\text{LocData}, \text{PubKey}) \quad (1)$$

$$\text{Dec}_M = \text{Decrypt}_{\text{EC P-224}}(\text{Enc}_M, \text{PrivKey}) \quad (2)$$

Here, Enc$_M$ represents the encrypted message, and Decrypt$_{\text{EC P-224}}$ and Encrypt$_{\text{EC P-224}}$ denote the decryption and encryption functions using the EC P-224 elliptic curve cryptography. LocData refers to the location data, PubKey is the public key used for encryption, and PrivKey is the private key used for decryption. This mechanism ensures that the location data remains secure and private, accessible only to the intended owner of the AirTag.
- **Samsung SmartTags**: Samsung SmartTags employ symmetric key cryptography, specifically AES in Cipher Block Chaining (CBC) mode with PKCS7 padding, for generating digital signatures within the device itself. This process is depicted in Equation (3), highlighting the sophisticated encryption mechanism utilized by SmartTags:

$$\text{Signature} = \text{AES}_{\text{CBC}}(\text{PKCS7}(\text{Data}), \text{Key}) \quad (3)$$

In this equation, Signature is the resultant digital signature generated by the SmartTag. The term AES$_{\text{CBC}}$ signifies the utilization of the Advanced Encryption Standard (AES) algorithm in Cipher Block Chaining (CBC) mode. The expression PKCS7(Data) represents the application of PKCS7 padding to the data prior to its encryption. The variable Key refers to the symmetric encryption key used in this process. This cryptographic approach, while enhancing the security of transmitted data, raises potential concerns regarding user privacy due to the centralized verification of signatures at the server side.

*2) Beacon Broadcast*
- **Apple AirTags**: Include a daily-changing public key and a frequently updating crypto counter for device anonymity.
- **Samsung SmartTags**: Feature a Privacy ID and a digital signature for unique identification and payload integrity.

*3) Lost Mode Activation*
Both trackers activate a "lost mode" when out of range, differing in beacon content (Tables I and II).

To facilitate a clearer understanding, we employ flowcharts, as shown in Figure 2 for Apple AirTags and Figure 3 for Samsung SmartTags. These diagrams effectively illustrate the sequence of operational steps for each type of tracker, starting from the activation of lost mode to the point where the owner receives the location data.

TABLE I: Detailed Breakdown of Apple AirTag Advertisement Data

| Byte # | Value | Description |
|---|---|---|
| 0 | 0x1E | Advertising data length: 31 (the maximum allowed) |
| 1 | 0xFF | Advertising data type: Manufacturer Specific Data |
| 2-3 | 0x004C | Apple's company identifier |
| 4 | 0x12 | Apple payload type to indicate a FindMy network broadcast |
| 5 | 0x19 | Apple payload length (31 - 6 = 25 = 0x19) |
| 6 | 0x10 | Status byte |
| 7-29 | Varies | EC P-224 public key used by FindMy network. Changes daily |
| 30 | 0-3 | Upper 2 bits of first byte of ECC public key |
| 31 | Varies | Crypto counter value; Changes every 15 minutes to a random value |

TABLE II: Detailed Breakdown of Samsung Galaxy SmartTag Advertisement Data

| Byte # | Value | Description |
|---|---|---|
| 0 | Varies | Tag state |
| 1-3 | Varies | Aging counter = (tagtime - 1593648000) / 900 |
| 4-11 | 0x4C | Unique identifier of a SmartTag |
| 12 | Varies | Region code, encryption flag, and UWB flag |
| 13-15 | 0x00 | Reserved |
| 16-19 | Varies | Digital signature |
| 20-31 | Unused | Unused |

*4) Comparison with Other Technologies*

The technological progression of BLE trackers is a remarkable reflection of advancements in both security and functional capabilities. Initially, these trackers were rudimentary, primarily serving as proximity alert systems. They functioned as basic locators, leveraging BLE technology to detect the proximity of tagged items within limited ranges. However, they were devoid of sophisticated features like advanced encryption algorithms or 'lost mode' functionalities, which are hallmarks of contemporary trackers.

In stark contrast, modern iterations of BLE trackers, notably

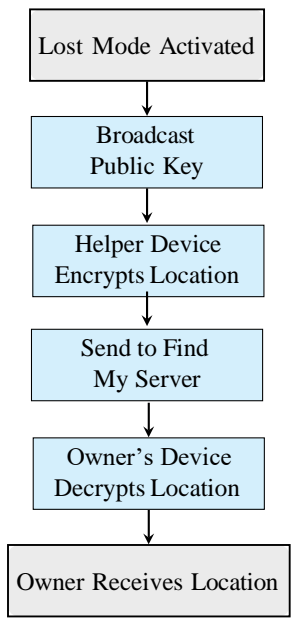

Fig. 2: Apple AirTag Operational Flowchart

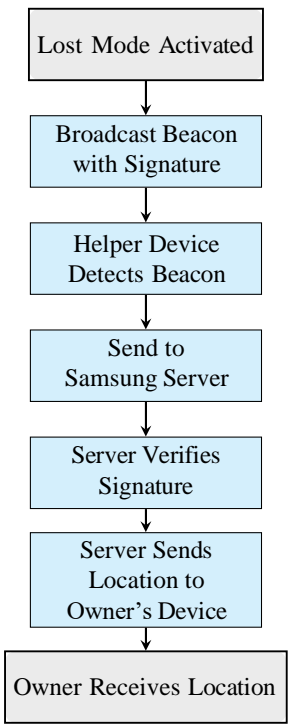

Fig. 3: Samsung SmartTag Operational Flowchart

Apple AirTags and Samsung SmartTags, have significantly eclipsed these early constraints. Exhibiting considerable enhancements in their security protocols and operational capabilities, these devices exemplify the forefront of BLE tracker technology. Apple AirTags utilize a public-key cryptography system (EC P-224), while Samsung SmartTags opt for symmetric key cryptography (AES/CBC/PKCS7), demonstrating a profound shift from the simplistic security mechanisms of their predecessors.

The introduction of 'lost mode' in these devices is a notable advancement. It enables the trackers to communicate their location beyond the immediate vicinity of the owner's device by tapping into a broader network of devices, reflecting not only technological innovation but also the expansion of the ecosystem in which these trackers operate.

Further, Apple AirTags and Samsung SmartTags are seamlessly integrated into larger technological frameworks, enhancing both their utility and user experience. AirTags are incorporated into Apple's extensive FindMy network, and SmartTags form part of Samsung's SmartThings network. This integration fosters a more unified and effective user experience, surpassing the capabilities of standalone trackers.

Compared to other market competitors, such as Tile or Chipolo, AirTags and SmartTags stand out with their advanced security features and seamless ecosystem integration. This synergy offers enhanced functionalities like more accurate location tracking and improved security, features that are typically absent in other BLE trackers.

## III. RELATED WORK

The literature review investigates the security and privacy considerations of BLE trackers from Samsung and Apple,

Offline Finding networks, continuity protocols, and other products utilizing Bluetooth technology. Heinrich et al. [5] identified two design and implementation vulnerabilities not accounted for in Apple's threat model. These vulnerabilities could potentially lead to location correlation attacks and unauthorized access to location histories from the past week. The researchers also reverse-engineered the FindMy protocols, demonstrating the feasibility of creating custom tracking devices using the OpenHaystack framework [6] that could leverage the FindMy network.

Yu et al. [7] explored Samsung's Offline Finding protocol within the Find My Mobile system. Using BLE, it locates Samsung devices and addresses privacy concerns from owner, finder, and vendor perspectives, covering device identification, malicious tracking, de-anonymization, and integrity issues. Privacy risks were identified due to protocol design and implementation. In a previous version of the Samsung Find My Mobile app, Char49 researchers [8] found multiple vulnerabilities enabling malicious apps to manipulate the app's URL endpoint, allowing unauthorized access to unprotected broadcast receivers.

Both Apple AirTags and Samsung SmartTags use the nRF52 series of System on Chips, commonly found in IoT devices supporting Bluetooth LE and Bluetooth Mesh. However, these chips are susceptible to power glitching attacks. Security analysis of AirTags revealed vulnerabilities in hardware and firmware. The AirTags' main firmware was extracted through voltage glitching attacks on the nRF chip, enabling activities like cloning, customizing soundsets, and using the accelerometer as a makeshift microphone [9].

An analysis of Apple's Continuity Protocols, facilitating data sharing among Apple devices, exposed vulnerabilities in BLE usage [10]. Researchers identified issues such as identity address exposure, infrequent MAC address randomization, and predictable incremental sequences in HandOff messages. These vulnerabilities ranged from minor, like battery level disclosure, to critical, such as plaintext phone number exposure, undermining BLE anti-tracking measures and enabling long-term passive tracking. In 2021, Mayberry et al. [11] examined Apple's anti-tracking measures and developed techniques to circumvent them. Bit Flipping manipulates the FindMy device type byte, avoiding tracker detection. The other techniques rely on frequent key rotations to thwart anti-tracking algorithms. The study revealed that Apple's iOS tracking detection fails to identify trackers using these techniques, allowing adversaries to track targets without detection.

BLE trackers have raised significant privacy concerns, as they have the potential to be utilized for tracking individuals. This could involve placing a BLE tracker on someone's vehicles or belongings, or even tracking the owner by collecting device information in public areas. Several papers studied the privacy implications of these trackers for Airtags [12]–[15] and SmartTags [16], [17]. However, it's important to note that privacy is not a primary focus of this paper; rather, the main emphasis is on security-related issues associated with BLE trackers.

Our research stands out by comprehensively encompassing all potential types of attacks on BLE tracking systems. It serves as a singular resource covering all security-related aspects of BLE trackers. Additionally, our study explores the inherent design challenges in developing such systems, introduces and assesses novel attack scenarios, and scrutinizes the impact of varied implementations in BLE tracking systems, with a particular focus on those developed by major vendors, Apple and Samsung. Furthermore, our work delves into the future trajectory of AirTag trackers, evaluating its potential implications for security

## IV. Attack Surface Analysis

In this section, we dissect the attack surface, considering the components of BLE tracking systems and the associated cryptographic framework.

### A. Attacks on the BLE tracker

1) **Physical Tampering:** Attackers could physically tamper with the BLE tracker to disable its functionality or modify it to perform malicious actions. There have been attempts to dump the firmware for Apple AirTags. In 2021, a group of researchers at the Technical University of Munich published a paper detailing their attempts to dump the firmware for Apple AirTags. They were able to extract some of the firmware, but they were not able to extract the master key [18]. They were able to reprogram the AirTag to generate a non-Apple URL while in lost mode, a vulnerability that an advanced attacker potentially could exploit to get high-value targets to open a custom malware site. On the other hand, there were attempts to dump the firmware of the SmartTag. Whid-injector [19] was able to dump the SmartTag firmware without further analysis. The ultimate goal of such physical tampering would be extracting the master key, which would allow cloning the AirTag. No successful attempts yet. Further voltage glitching was employed to attack on the nRF chip, researchers were able to extract and reverse engineer the main firmware. They further manipulate the firmware by adding functionality [9], altering, and facilitating actions such as playing sound sequences and downgrading the nRF and U1 firmware.

2) **Firmware Exploitation:** In cases where BLE tracker firmware lacks proper security measures, attackers could potentially exploit weaknesses, leading to unauthorized access or control of the tracker. For instance, there is a stored XSS vulnerability affecting the data generated on the found.apple.com domain for each AirTag, specifically within the phone number field where a malicious payload can be inserted [20]. Notably, it's essential to highlight that AirTag lacks secure boot and permits the execution of altered firmware, as pointed out in [21]. Consequently, there is potential for crafting firmware that can manipulate the system, as discussed in [22]. This makes it feasible to create attacks on the tracking system using custom firmware. For example, researchers were able to modify the firmware to prevent it from sending out privacy warnings [9].

3) Signal Spoofing: Attackers have the potential to replicate the Bluetooth signal of a genuine tracker, deceiving its owner. A Raspberry Pi proves to be a great tool for broadcasting BLE beacons. Developers can create applications for transmitting these signals by leveraging the Pi's built-in Bluetooth capabilities or incorporating a compatible BLE USB dongle. In our research, we utilized a BLE-enabled Pico Pi [23] to develop a program for custom beacon broadcasting. This program was employed to broadcast a token resembling that of an AirTag in lost mode [24]. We conducted an experiment where both the genuine AirTag and the Pico Pi, with the spoofed token, broadcasted simultaneously, even though they were hundreds of miles apart. Notably, as shown in Figure 4, the spoofed location initially appeared on the Find My App before bouncing back to the authentic AirTag's location.

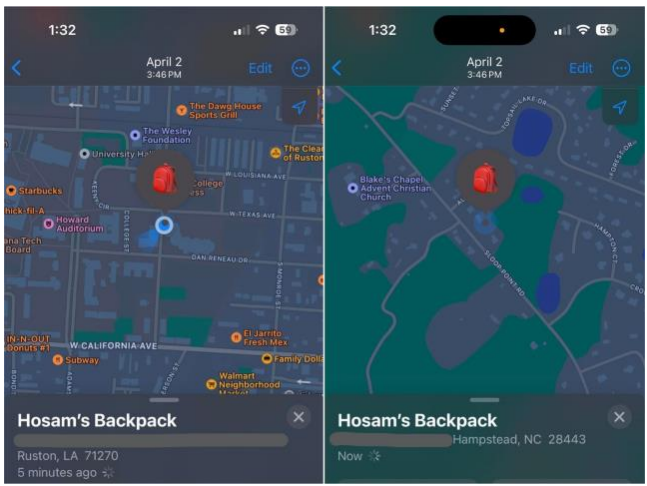

Fig. 4: Beacon spoofing attack on Apple's Find My.

This attack was also carried out successfully on the SmartTag. However, it is important to point out that the whole beacon for SmartTag changes every 15. In contrast, the AirTag changes only a 1-byte every 15 minutes. Despite this alteration, it still permits the execution of the attack by broadcasting all 256 possibilities.

4) Eavesdropping: BLE communication between the tracker and smartphones might be susceptible to eavesdropping attacks, enabling the theft of pairing messages for potential exploitation in replay or relay attacks. While the logistics of these attacks can be challenging, requiring continuous sniffing during the pairing process, it remains possible for an attacker to pose as the legitimate device and pair itself instead or alongside the authentic device.

5) Jamming: Jamming attacks on BLE trackers involve deliberately disrupting their communication with either the helper device or the owner's device. These attacks pose a significant threat to devices like AirTags because they target the renewal of the P-224 key, a crucial security element of AirTags systems. Generally, jamming attacks relatively easy to execute. BLE operates in the 2.4 GHz frequency band, a general use frequency band. Attackers can employ affordable hardware like RF transmitters or software-defined radios to flood BLE channels with noise, effectively drowning out legitimate communication. This leads to a denial-of-service situation where BLE devices can't function as intended. Due to their low-power nature, BLE devices prioritize power efficiency, making them vulnerable to disruption. Their inability to boost transmission power effectively leaves them susceptible to jamming. In experiments, jamming was effortlessly accomplished using devices like the Flipper Zero [25] and Blade-RF [26]. This disrupted communication between the owner device and the tracker, activateing the lost mode and interfering with the helper device's ability to receive location updates. Both AirTags and SmartTags face an equal vulnerability to jamming attacks. One potential application of these attacks is the creation of no-tracking zones, rendering BLE trackers unusable. This strategy can be implemented in high-security areas to thwart the tracking of item movement within those designated zones.

B. Attacks on Helper devices:

1) App Security: If the smartphone application linked to the BLE tracker system has security weaknesses, malicious actors may exploit these flaws to access users' location data. If an application has sufficient access to the BLE interface, an auxiliary device can be manipulated to gather BLE beacons and transmit them to an external entity. This information could then be exploited to track the location of these BLE devices or potentially launch a large-scale attack on the BLE tracking system, as elaborated in the subsequent section. It's important to note that smartphone operating systems implement safeguards against such threats, mandating user approval for granting applications access to Bluetooth functionality [27], [28]. Nevertheless, it's essential to highlight that this type of access is considered when the app is in an active state. Different considerations come into play when dealing with background scanning. iOS does not support general, continuous background processes for various purposes [29]. In contrast, there was a time when conducting background scans for beacons on Android was relatively straightforward – you could establish a service using standard Bluetooth Application Programming Interface (API)s. However, as the years have passed, Google has significantly restricted the ability to execute dependable background processing, as described in [30]. Nonetheless, it's worth acknowledging that such attacks may still be viable on compromised smartphone operating systems and through compromised applications.

2) Bluetooth Security: Smartphone Bluetooth connections might be vulnerable to exploitation, allowing attackers to intercept or manipulate data between the smartphone and the tracker. In BLE, pairing is a secure process. Encryption in BLE relies on the 128-bit Advanced Encryption

Standard — Counter with CBC-MAC. This algorithm utilizes a Long-Term Key to generate a 128-bit "shared secret" key. BLE ensures authentication by digitally signing the data with the Connection Signature Resolving Key (CSRK) . When sending data, the transmitting device appends a signature after the Data Protocol Data Unit, which is then verified by the recipient using the CSRK. However, there has been attacks on BLE pairing process such as downgrade attacks [31], injection attacks [32], and key renegotiating attacks [33]. Such attacks can possibly used to hijack the pairing and steal the secret key used to generate beacons.

3) Location spoofing: A falsified location on the helper device can be employed to relay inaccurate location information to the cloud. In this instance, the location is manipulated by a helper device reporting a beacon of a BLE tracker but with a spoofed location. This attack leads to find-my-device technologies reporting the location of a helper device that does not correspond to the actual location of a BLE tracker. To examine this attack scenario, we perform GPS location spoofing on a helper device, with the objective of evaluating the ability of Apple and Samsung servers to detect manipulated location data. This experiment was executed on both an iPhone and a Samsung device as outlined below:

   a) iPhone: To manipulate the location on an iPhone, the tool employed was Tenorshare iAnyGo. iAnyGo is a multifaceted iOS GPS location spoofing tool crafted to utilize developer mode. Location spoofing is permitted on iPhones in developer mode to facilitate the testing of location-based applications while safeguarding user privacy. In this trial, iAnyGo was utilized to falsify the location on a helper device. The helper device was positioned within the range of an AirTag in lost mode, while simultaneously monitoring the reported location on the owner device. The iPhone did not report the airtag location to the cloud, possibly because safeguards are in place to prevent iPhones from reporting location while in developer mode. Testing this scenario with a jailbroken version of iOS was not feasible, considering the potential failure of Apple services to run on jailbroken iPhones [34].
   b) Samsung: In this experiment, a rooted Samsung phone was employed, utilizing the GPS Joystick app [35] to manipulate the device's GPS location through various modes. Due to Samsung's validation using IP addresses, the attack was conducted in close proximity, with the location of the spoofed devices being less than 1 mile from the original. For greater distances, a Virtual Private Network (VPN) could be implemented to change the IP and match the geo-location. The System Mode, leveraging system privileges to simulate location, was activated on the helper device. Subsequently, the Galaxy tag was positioned in proximity to the manipulating helper device, as can be seen in Figure 5 the spoofed location was reported to the Samsung SmartThings network, which was observed by the owner device.

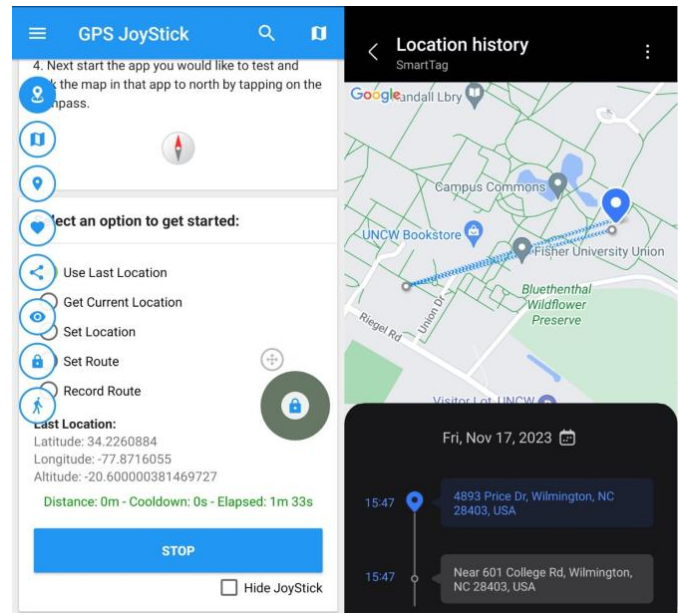

Fig. 5: GPS joystick to report spoofed locaiton to Samsung server

## C. Attacks on owner device:

In this case, the smartphone is attacked by any elements in the system. One big attack surface in Apple FindMy framework is that the it can be used as a side channel to send arbitrary data to the owner device. Positive Security researchers [36] found a way to exploit Apple's Find My network, 'Send My,' which lets arbitrary data be transmitted globally, not just device locations. They demonstrated the risk by integrating a keylogger into a USB keyboard, using Bluetooth to relay sensitive data via Find My, making detection difficult. Arbitrary data can be crafted to include malicious code or malware. When the smartphone processes this data, it can execute the embedded code, potentially taking control of the device, stealing data, or conducting other malicious activities.

## D. Attacks on the Cloud

1) Data Breach: Insufficiently securing the cloud server responsible for storing location data can potentially grant unauthorized access to the stored data, endangering both user privacy and system security. The incidence of data breaches is on the rise [37]. Notably, several location servers, including those of Sprint [38], Fitbit [39], and Garmin [40], have been compromised. In 2020, Samsung experienced a breach in the "Find My Mobile" app, which came pre-installed on stock Samsung Android distributions. Conversely, Apple's different implementation of location tracking, as observed with AirTag, enhances its resilience against breaches. This is because AirTags require location data to be decrypted solely on the owner's device. In contrast, Samsung's approach allows the service operator (Samsung) to access the reported locations of any SmartTag devices and to associate these locations with the device owner using sensitive data stored on their server, such as encryption keys and privacy configurations. Furthermore, helper devices are susceptible to tracking, as each location report includes a unique authentication token issued by the location server, which can be linked to a specific helper device for an extended duration [16]. Consequently, a data breach would have a more pronounced impact on the security of the Samsung system.

2) Denial of Service Attacks: In the realm of cybersecurity, malevolent actors may attempt to flood a cloud server with an excessive illegitimate requests, a tactic that leads to service disruptions and obstructs legitimate users from accessing the system. These attacks involve user devices directing requests toward the cloud servers. While cloud servers may boast defenses against conventional HTTP or TCP-based flood attacks, they can still be vulnerable to the spoofing attacks. A spoofing attack unfolds when a helper device forwards beacons to the cloud, necessitating processing and validation of these beacons, which in turn incurs computational expenses prior to confirming the invalidity of the beacon. The magnitude of this cost varies according to the specific implementation. In the case of SmartTag, privacy IDs are deployed to distinguish unique SmartTags, using a symmetric encryption secret generator, as detailed in [16]. Consequently, the server generates an identical secret and conducts a comparison. In contrast, in the context of Apple AirTags, the beacons are forwarded to the owner device, which bears the weight of the beacon verification.

3) API Vulnerabilities: In the context of cloud services, when an API is made available for communication with other software applications, it introduces potential security risks related to the API itself. These risks encompass threats like injection attacks, where malicious code can be inserted, and authentication bypass, which can circumvent the security measures. At the time this paper was authored, neither AirTag nor SmartTag had integrated support for an API, meaning they did not expose such a programming interface for external applications to interact with. However, it's worth mentioning that Apple introduced a feature called the "Nearby Interaction API" as part of iOS 15. This API empowers U1-enabled iOS devices to perform tasks such as determining the distance, orientation, and proximity to UWB-enabled tags or beacons, functioning in a manner similar to AirTags [41]

4) Large-Scale Botnet Spoofing Attacks: Here we discuss an attack scenario concerning BLE trackers, specially AirTags as they do not authenticate the identity of AirTags before accepting location updates. The attack we are discussing extends the concept of signal spoofing attacks previously described. This particular attack leverages a botnet comprised of BLE-enabled devices. It's worth noting that a recent study by Statista has revealed that there are approximately 6.6 billion IoT devices in the United States as of 2023. [42]. Many of IoT devices are becoming BLE enabled. According to a report by MarketsandMarkets, BLE is expected to be the fastest-growing connectivity technology for IoT devices in the coming years. By 2028, BLE is expected to account for over 30% of all IoT devices [43]. If an attacker has compromised a number of BLE-enabled IoT devices. He/She can launch a mass spoof campaign in which bots broadcast a possible spoofed tags. Apple's AirTag, which had 55 million units sold by the end of 2022 according to Yahoo Finance [44], is more susceptible to this type of attack. In the AirTag's "Lost Mode," the entire P-224 value changes only once every 24 hours, and the last two bytes of this value change every 15 minutes. In contrast, the BLE interface has the capacity to transmit a significant number of beacons every second. Our experiments demonstrated the feasibility of programming a Raspberry Pi Pico to broadcast 10 distinct beacons per second. As a result, such an attack can effectively target and render a single AirTag or multiple AirTags untraceable. This attack goes as follows:

  a) The attacker determine the target(s). A lost mode beacon of the of the target AirTag is needed.
  b) The attacker sends orders to the BLE enabled botnet to broadcast the beacon in (A) with different values of the variable counter ( The last byte of the beacon). The attacker can use a single device to broadcast multiple token by rotating beacons with different counter number on the same BLE interface.
  c) Helper devices will pick up the tokens and report them to the cloud server. Multiple location will be fetched by the phone without the phone knowing which location is legitimate.

### E. Attacks on Cryptography

Apple utilizes the NIST P-224 elliptic curve for Find My™ technology. This includes iPhones, iPads, Mac devices, and AirTags. The P-224 curve is a specific curve recommended by NIST on a specific prime field $\mathsf{F}_p$ with some additional parameters and random seeds to specify its creation. A curve $E$ over a prime field is defined in Equation (4)

$$E : y^2 \equiv x^3 - 3x + b \pmod{p} \quad (4)$$

This equation forms the basis for a pseudo-random curve for prime $p$ of order $n$ where $p, n$ are given by NIST. NIST also provides a seed for pseudo-random number generators, the output of a SHA-1 based algorithm $c$, the constant coefficient $b$, and the base points' $x, y$ coordinates $G_x, G_y$. The constant coefficient $b$ must satisfy $b^2 c \equiv -27 \pmod{p}$.

FindMy devices generate P-224 private encryption key pairs, denoted $d, P$, where $d$ is the private key and $P$ is the public key. The device also generates a 256-bit secret $SK_0$ and a counter *i* initialized at 0. These private key pairs are synced only on the users devices with end-to-end encryption using iCloud Keychain. The addition secret symmetric key $SK_0$ and the counter *i* are used to create ephemeral keys $d_i$, $P_i$ based on the following recursions:

$$SK_i = KDF(SK_{i-1}, "update") \quad (5)$$

$$(u_i, v_i) = KDF(SK_i, "diversify") \quad (6)$$

$$d_i = u_i d + v_i \quad (7)$$

$$P_i = u_i P + v_i G \quad (8)$$

Where KDF is a key derivation function. The subsequent pairs $d_i$, $P_i$ are used in the device beacon for transmitting location data in a secure manner where the holder of $d$, $P$, $SK_0$ can generate the current symmetric key $SK_i$, and decrypt the message and obtain the device location. This forms an elliptic curve Diffe-Hellman ephemeral (ECDHE) scheme where the account owner can apply the inverse operations to decrypt location reports. The details on Apples methods are detailed in an Apple support page post on Platform Security [45]. This system of offline finding can be elaborated on in the context of attacks on cryptography by an adversary model. A useful model from [46] specifies local application (**A1**), proximity-based (**A2**), network-based (**A3**), and service operator (**A4**) adversaries.

Since Apple devices utilize NIST P-224 for their lost mode beacons, their security is restricted to at least that of the encryption implementation. For an **A1** attacker capable of skimming beacon signals, the privacy of location history relies solely on this encryption implementation. Similarly, an **A3** attacker may be able to capture and aggregate location reports which will again rely on the security of the encryption scheme. Finally, a service provider or breach of the service provider will assume an **A4** attacker who has access to location reports of many users whose location data privacy will again depend at least on the encryption scheme.

For these reasons, there are yet more incentives to analyze the security of the NIST P-224 encryption scheme, along with the symmetric key scheme implemented by Apple. The website SafeCurves [47] provides a list of potential weaknesses of the NIST P-224 curve. One such weakness is *twist attacks* which for certain implementations of P-224 put the number of operations required to break the encryption of P-224 ECDHE on the order of $2^58.4$ operations which is achievable. An example of this attack, labeled a *Fault Attack* is from Fouque et al. [48].

Heinrich et al. [46] elaborate on potential vulnerabilities from **A1**, **A3** and **A4** based attackers, leading to the potential for unauthorized access of location history, spoofing of device locations, and correlation of user locations, respectively. In addition to these potential vulnerabilities, [46] notes that there is also the issue of insecure key storage, leading to a greater risk for private location data in the case that an attacker had combined capabilities of **A1** and **A2** or **A1** and **A4** or a combination of the three.

These types of attacks and vulnerabilities are not exclusive to Apple products. Similar to the setting of an **A1** attacker for Apple devices, several hardcoded keys are at risk of extraction on Samsung devices as detailed by [8] leading to potential data leakage and privacy risks. Another concern lies in the compact signature size within the beacon. With the signature comprising only 4 bytes, there is an elevated risk of collusion [49].

## V. Discussion

In this section, we explore various aspects pertaining to BLE trackers, covering design challenges, the impact of different implementations, and future trends. Each aspect is discussed in its dedicated subsection.

### A. Design Challenges

BLE, which stands for Bluetooth Low Energy, indicates a focus on low energy consumption in devices. This characteristic significantly impacts various aspects of design. Firstly, design limitations emerge due to considerations for battery life, leading to challenges in reducing power consumption. This results in compromises, such as using weaker cryptographic functions [50] and limiting the frequency of changing advertisements [51]. Another power-related issue involves restricting the range of BLE trackers, necessitating a larger network of helper devices for efficient tracking. Additionally, BLE trackers are vulnerable to interference and obstacles. Operating on the widely used 2.4 GHz frequency, BLE signals can be affected by interference from other wireless devices or physical barriers like walls, reducing the accuracy of location tracking.

Firmware and software updates pose another significant challenge. The process of updating firmware or software on BLE trackers can be complex, particularly when it requires user interaction. Notably, both AirTag and SmartTag were not designed to receive firmware updates, making it challenging to patch these systems in case of security issues. Finally, the absence of secure boot processes in both devices highlights the challenges in creating secure yet efficient BLE trackers. The exclusion of secure boot processes, driven primarily by cost concerns and the fear of device bricking, introduces a notable security risk. As BLE trackers advance, the demand for more robust security measures becomes crucial, especially with their growing integration into broader IoT ecosystems and the incorporation of features like health monitoring. An additional design challenge pertains to cost considerations; the devices must be economically feasible to penetrate the mass consumer market. This imposes limitations on the utilization of specialized power-saving chips or unique battery types.

### B. Different Design Impact

The development of BLE trackers, such as Apple AirTags and Samsung SmartTags, involves a careful balancing act between functionality and security, especially given the inherent constraints of BLE technology mentioned earlier. The

931-byte limit imposed on the size of beacon data that can be transmitted by BLE devices necessitates a strategic approach to implementing security measures. The differing designs of AirTags and SmartTags highlight distinct strategies for addressing the challenges arising from these constraints. Apple's strong emphasis on user privacy is evident in its design, which ensures that only the owner device possesses the private key necessary to decrypt the location of lost AirTags, encrypted in the P-224 public key. However, this privacy-centric approach entails certain trade-offs, particularly in the area of device authentication. As discussed in previous sections, the vulnerability of AirTags to device spoofing is a notable concern, stemming from the absence of beacon authentication. This highlights a weakness in Apple's security strategy, indicating that the prioritization of privacy may have resulted in oversights in device authentication measures.

On the other hand, Samsung's design places a strong emphasis on ensuring the authenticity of beacons through the use of digital signatures. This approach is effective in countering spoofing concerns, providing a layer of security that addresses one of the potential vulnerabilities identified in Apple's AirTags. However, this focus on beacon authenticity introduces its own set of challenges, particularly regarding the potential for privacy breaches. The concern here is that if the cloud infrastructure supporting the digital signatures is compromised, it could lead to unauthorized access or tracking, undermining the user's privacy. The analyzed designs illustrate the complex interplay between functionality, security, and privacy. It highlights that the choices made in addressing the limitations of resource-constrained environments can have significant implications for the overall security and privacy of these devices. As technology continues to advance, it becomes increasingly crucial for developers to strike the right balance between innovation, user convenience, and robust security measures

## C. Future Trends

Looking forward, the landscape of BLE technology is poised for significant advancements that will invariably influence security considerations. We anticipate several key trends and developments:

*1) Advanced Encryption Techniques*

Future BLE trackers are likely to adopt more sophisticated encryption methods. Implementing advanced cryptographic frameworks, such as those suggested in [52], can provide stronger security against emerging threats.

*2) Improved Battery Technology*

The advent of new battery technologies could alleviate current limitations, enabling more complex security features without compromising device longevity.

*3) Interoperability and Ecosystem Integration*

Enhanced interoperability between different BLE devices and integration into larger ecosystems will necessitate comprehensive security evaluations. This integration, while beneficial for functionality, introduces new vulnerabilities that must be addressed.

*4) Secure Boot and Anti-Spoofing Technologies*

Implementing secure boot processes and advanced anti-spoofing measures will be crucial in safeguarding against firmware tampering and spoofing attacks.

*5) Regulatory Compliance and Privacy*

As BLE trackers become more prevalent, they will face increased scrutiny under privacy regulations. Manufacturers will need to ensure compliance with evolving data protection laws.

*6) Zero-Trust Security Model*

In an interconnected ecosystem, adopting a zero-trust security model will be essential. This approach involves verifying every device and transaction, regardless of the network, to mitigate risks associated with expanded connectivity.

*7) Machine Learning and AI Integration*

The integration of AI and machine learning algorithms can provide predictive security measures, identifying and mitigating potential threats proactively.

*8) User Awareness and Education*

With the progression of technology, the significance of educating users on best practices for privacy and security is on the rise. This entails raising awareness about potential privacy concerns associated with the use of BLE trackers, as well as informing users about potential vulnerabilities and providing instructions on how to maintain the security of their devices.

While current BLE tracker designs offer a balance between functionality and security, future advancements in BLE technology present both opportunities and challenges. Embracing these advancements will require a holistic approach to security, considering not only the technical aspects but also user behavior, regulatory compliance, and the broader ecosystem in which these devices operate. The continual evolution of BLE technology will thus play a pivotal role in shaping the security landscape of these devices.

## VI. CONCLUSION

In this comprehensive study, we have undertaken a thorough examination of the security vulnerabilities inherent in BLE tracking systems, specifically focusing on two prominent examples: Apple AirTags and Samsung SmartTags, and their associated cryptographic frameworks. Our investigation delves into a range of potential attack vectors, spanning from physical tampering and firmware exploitation to signal spoofing, eavesdropping, jamming, app security vulnerabilities, Bluetooth security weaknesses, location spoofing, as well as threats directed at owner devices and cloud-based components.Our analysis underscores that while BLE tracking systems offer substantial advantages in terms of finding lost items and enhancing user convenience, they simultaneously introduce security risks that necessitate vigilant attention. Apple's design emphasizes user privacy by minimizing intermediaries, yet this emphasis may compromise device authentication, as our successful spoofing of AirTags demonstrates. In contrast, Samsung SmartTags prioritize the prevention of beacon spoofing but raise concerns regarding cloud infrastructure security and the potential for privacy breaches. Developing these tracking devices is further complicated by their constrained resources, primarily tailored for conserving battery life. The incorporation of secure boot processes and advanced security features may entail increased manufacturing costs but is indispensable in mitigating risks and upholding the integrity of BLE tracking systems. Anticipating forthcoming iterations of these devices, manufacturers are anticipated to enhance security features as they become more intricately integrated into the expanding IoT landscape, subject to heightened scrutiny related to privacy regulations. In conclusion, while BLE tracking systems hold immense promise, their security must evolve to counteract the myriad attack vectors detailed in this study. This paper lays a solid foundation for comprehending the security landscape of BLE trackers, underscoring the necessity for ongoing research and development to fortify the safeguarding of user data and privacy in the ever-evolving realm of IoT and interconnected devices.

## REFERENCES

[1] D. Hortelano, T. Olivares, M. C. Ruiz, C. Garrido-Hidalgo, and V. López, "From sensor networks to internet of things. bluetooth low energy, a standard for this evolution," *Sensors*, vol. 17, no. 2, p. 372, 2017.
[2] Apple, "Find my," 2023.
[3] Samsung, "Smartthings," 2023.
[4] Flaticon, "Flaticon: Free vector icons," 2023.
[5] A. Heinrich, M. Stute, T. Kornhuber, and M. Hollick, "Who can find my devices? security and privacy of apple's crowd-sourced bluetooth location tracking system," *arXiv preprint arXiv:2103.02282*, 2021.
[6] seemoo-lab, "OpenHaystack GitHub repository," 2023.
[7] T. Yu, J. Henderson, A. Tiu, and T. Haines, "Privacy analysis of samsung's crowd-sourced bluetooth location tracking system," *arXiv preprint arXiv:2210.14702*, 2022.
[8] Char49, "Samsung Find My Mobile vulnerability," 2019. Accessed: 01-Oct-2022.
[9] F. Roth and K. Freyer, "Airtag of the clones: Shenanigans with liberated item finders," 2022.
[10] J. Martin, D. Alpuche, K. Bodeman, L. Brown, E. Fenske, L. Foppe, T. Mayberry, E. C. Rye, B. Sipes, and S. Teplov, "Handoff all your privacy - a review of apple's bluetooth low energy continuity protocol," *Proceedings on Privacy Enhancing Technologies*, vol. 4, pp. 34–53, 2019.
[11] T. Mayberry, E. Fenske, D. Brown, J. Martin, C. Fossaceca, E. C. Rye, S. Teplov, and L. Foppe, "Who tracks the trackers? circumventing apple's anti-tracking alerts in the find my network," in *Proceedings of the 20th Workshop on Workshop on Privacy in the Electronic Society*, pp. 181–186, 2021.
[12] J. Briggs and C. Geeng, "Ble-doubt: Smartphone-based detection of malicious bluetooth trackers," in *2022 IEEE Security and Privacy Workshops (SPW)*, pp. 208–214, 2022.
[13] K. O. E. Müller, L. Bienz, B. Rodrigues, C. Feng, and B. Stiller, "Homescout: Anti-stalking mobile app for bluetooth low energy devices," in *2023 IEEE 48th Conference on Local Computer Networks (LCN)*, pp. 1–9, 2023.
[14] N. Shafqat, N. Gerzon, M. Van Nortwick, V. Sun, A. Mislove, and A. Ranganathan, "Track you: A deep dive into safety alerts for apple airtags," *Proceedings on Privacy Enhancing Technologies*, vol. 4, pp. 132–148, 2023.
[15] N. McBride, "Apple airtags as people trackers," in *Tracking People*, pp. 221–238, Routledge, 2023.
[16] T. Yu, J. Henderson, A. Tiu, and T. Haines, "Privacy analysis of samsung's crowd-sourced bluetooth location tracking system," 10 2022.
[17] T. Despres, N. Davis, P. Dutta, and D. Wagner, "Detagtive: Linking macs to protect against malicious ble trackers," in *Proceedings of the Second Workshop on Situating Network Infrastructure with People, Practices, and Beyond*, pp. 1–7, 2023.
[18] D. Wegemer, M. Wieser, K. Rieck, and F. Buchholz, "Extracting airtag firmware: A brief overview," *arXiv preprint arXiv:2107.02182*, 2021.
[19] whid injector, "Samsung smarttag hack," 2023.
[20] E. Kovacs, "Hackers can exploit apple airtag vulnerability to lure users to malicious sites," *SecurityWeek*, 2022.
[21] A. Catley, "Apple AirTag Reverse Engineering," 2023.
[22] ZDNet, "I built a custom AirTag that Apple will hate me for (and how you can do it, too)," 2023.
[23] R. Pi, "Raspberry Pi Pico," 2023.
[24] mCarlG, "picoTag," 2023.
[25] Anon, "How to: Jamming frequencies w/ the flipper zero," 2023.
[26] M. Tan, C. Wang, B. Xue, and J. Xu, "A novel deceptive jamming approach against frequency diverse array radar," *IEEE Sensors Journal*, vol. 21, no. 6, pp. 8323–8332, 2021.
[27] T. Baugher, "Bluetooth and location services permission tips," 2023.
[28] A. Developers, "Bluetooth permissions," 2023.
[29] A. Inc., "Accessory design guidelines for apple devices," 2023.
[30] BeaconZone, "Background scanning on android," 2023.
[31] Y. Zhang, J. Weng, R. Dey, Y. Jin, Z. Lin, and X. Fu, "Breaking secure pairing of bluetooth low energy using downgrade attacks," in *29th USENIX Security Symposium (USENIX Security 20)*, pp. 37–54, 2020.
[32] S. Pallavi and V. A. Narayanan, "An overview of practical attacks on ble based iot devices and their security," in *2019 5th International Conference on Advanced Computing & Communication Systems (ICACCS)*, pp. 694–698, 2019.
[33] A. C. Santos, J. L. S. Filho, Á. Í. Silva, V. Nigam, and I. E. Fonseca, "Ble injection-free attack: a novel attack on bluetooth low energy devices," *Journal of Ambient Intelligence and Humanized Computing*, pp. 1–11, 2019.
[34] Apple, "Unauthorized modification of ios," 2023.
[35] T. A. Ninjas, "Gps joystick faq," 2023.
[36] BleepingComputer, "Apple's find my network can be abused to steal keylogged passwords," *Bleeping Computer*, 2023.
[37] M. Kaur and A. B. Kaimal, "Analysis of cloud computing security challenges and threats for resolving data breach issues," in *2023 International Conference on Computer Communication and Informatics (ICCCI)*, pp. 1–6, 2023.
[38] T. Cymru, "Sprint location tracking vulnerability," 2015.
[39] H. I. S. Staff, "61m fitbit, apple users had data exposed in wearable device data breach," *Health IT Security*, 2021.
[40] Garmin, "Garmin services and production go down after ransomware attack," 2020.
[41] A. Inc., *Nearby Interaction*, 2023.
[42] Statista, "Iot connected devices worldwide 2019-2030."
[43] MarketsandMarkets, "Wireless connectivity market - global forecast to 2028," 2023.
[44] Yahoo Finance, "Apple AirTags Bluetooth trackers officially announced," *Yahoo Finance*.
[45] Apple, "Find my security," 2023.
[46] A. Heinrich, M. Stute, T. Kornhuber, and M. Hollick, "Who can find my devices? security and privacy of apple's crowd-sourced bluetooth location tracking system," *CoRR*, vol. abs/2103.02282, 2021.
[47] "Safecurves." https://safecurves.cr.yp.to/. Accessed: 2023-10-25.
[48] P.-A. Fouque, R. Lercier, D. Réal, and F. Valette, "Fault attack on elliptic curve montgomery ladder implementation," in *2008 5th Workshop on Fault Diagnosis and Tolerance in Cryptography*, pp. 92–98, 2008.
[49] F. Bao, "Colluding attacks to a payment protocol and two signature exchange schemes," in *International Conference on the Theory and Application of Cryptology and Information Security*, pp. 417–429, Springer, 2004.
[50] Z. Han, J. Katz, A. Menezes, and M. Strand, "Security analysis of the bluetooth low energy protocol," *ACM Transactions on Information and System Security (TISSEC)*, vol. 16, no. 3, pp. 1–54, 2013.
[51] M. Al-Qawasmeh, A. Ghebleh, N. Aslam, and T. Wan, "Energy-efficient bluetooth beaconing for proximity-based indoor positioning applications," *IEEE Access*, vol. 7, pp. 130622–130637, 2019.
[52] T. Mayberry, E.-O. Blass, and E. Fenske, "Blind my—an improved cryptographic protocol to prevent stalking in apple's find my network," *Proceedings on Privacy Enhancing Technologies*, vol. 1, pp. 85–97, 2023.